\documentclass[11pt]{article}
\usepackage{moriond,epsfig}

\def\Journal#1#2#3#4{{#1} {\bf #2}, #3 (#4)}

\def\PLB{{\em Phys. Lett.}  B}

\def\PRD{{\em Phys. Rev.} D}

\def\be{\begin{equation}}
\def\ee{\end{equation}}
\def\bea{\begin{eqnarray}}
\def\eea{\end{eqnarray}}

\newcommand{\TeV}{\ensuremath{\mathrm{Te\kern -0.1em V}}}
\newcommand{\GeV}{\ensuremath{\mathrm{Ge\kern -0.1em V}}}
\newcommand{\MeV}{\ensuremath{\mathrm{Me\kern -0.1em V}}}

\newcommand{\zll}{{ Z}^0\rightarrow{ \ell}^+ { \ell}^-}

\newcommand{\wlnu}{{ W^{\pm}} \rightarrow{ \ell}^{\pm}\nu} 

\newcommand{\zee}{{ Z}^0\rightarrow{ e}^+ { e}^-}
\newcommand{\zmm}{{ Z}^0\rightarrow{ \mu}^+ { \mu}^-}

\newcommand{\wenu}{{ W^{\pm}} \rightarrow{ e}^{\pm}\nu} 
\newcommand{\wmnu}{{ W^{\pm}} \rightarrow{ \mu}^{\pm}\nu} 
\newcommand{\wtnu}{{ W^{\pm}} \rightarrow{ \tau}^{\pm}\nu}

\newcommand{\pt}{\ensuremath{p_{\rm T}}}

\newcommand{\ppbar} {p\overline{p}}
\newcommand{\pbinv} {{\mathrm{pb}^{-1}}}
\newcommand{\fbinv} {{\mathrm{fb}^{-1}}}

\newcommand {\ra}     {\mbox{$~\rightarrow~$}}

\newcommand {\abseta} {\mbox{$\mid \eta \mid $}}

\newcommand{\met}{\mbox{${\hbox{$E$\kern-0.6em\lower-.1ex\hbox{/}}}_T\:$}}
\newcommand{\metx}{\mbox{${\hbox{$E$\kern-0.6em\lower-.1ex\hbox{/}}}_T\:$}}
\newcommand{\mety}{\mbox{${\hbox{$E$\kern-0.6em\lower-.1ex\hbox{/}}}_T\:$}}

\newcommand{\bit}{\begin{itemize}}
\newcommand{\eit}{\end{itemize}}
\newcommand{\bce}{\begin{center}}
\newcommand{\beq}{\begin{equation}}
\newcommand{\ece}{\end{center}}
\newcommand{\rg}{\rightarrow}

\begin{document}
%
\vspace*{4cm}
\title{ELECTROWEAK PHYSICS RESULTS AT CDF}

\author{GIULIA MANCA\footnote{manca@fnal.gov.}\\ (On behalf of the CDF collaboration)}

\address{University of Liverpool, Department of Physics,\\ 
Oxford street, Liverpool L69 7ZE, United Kingdom}

\maketitle\abstracts{
The Run II physics program of CDF is proceeding 
 with approximately 200 $\pbinv$ of analysis quality data collected at the 
center-of-mass energy of 1.96 \TeV. The Electroweak measurements are among the 
first and most important benchmarks towards the best understanding
of the detector and testing of the Standard Model.
We present precision measurements of the $W$ and $Z$ inclusive cross sections 
and decay asymmetries, 
recent results in di-boson physics and their sensitivity to
new physics, and preliminary studies
for the $W$ mass measurement.
}

\section{Introduction}

Studying the electroweak sector of the Standard Model(SM) is pivotal
in the CDF physics programme for Run II at the Tevatron.
The large number of $W$ bosons 
and the possibility to explore high mass $Z/\gamma^*$
are exceptional opportunities at hadron-colliders.
and some measurements at the Tevatron have unique sensitivities to the
ratio of the $u(\bar{u})$ to $d(\bar{d})$ parton distribution
functions in the proton(anti-proton).
The measurement of the $W$ mass, one of the
primary goals of the CDF electroweak programme, is 
central to constraining the mass of the SM Higgs boson
and thus to a deep and complete understanding of the 
Standard Model and its extensions. $W$ and $Z$ boson cross section 
measurements are 
key milestones in the understanding and calibration of
the detectors, and a starting point for any 
advanced measurement or discovery. 
Studies of boson asymmetries and di-boson production processes 
are based on clean and well-understood signatures which are
robust tests of the Standard Model and allow one to explore 
beyond-the-Standard-Model scenarios with the total integrated
luminosity of 4 to 8 $\fbinv$ expected in this decade.
%
Both the accelerator~\cite{peter} and the CDF detector have undergone
major upgrades in order to handle the increase in luminosity 
and energy achieved during Run II.
The increase in center-of-mass energy from 1.8 to 1.96 \TeV\
 is directly mirrored 
in an increase of the production cross sections, 
which is approximately 10\% for the $W,Z$ and 
in the case of the top is about 30\%~\cite{top,ken}.
The upgraded CDF detector has a completely 
new tracking system, extended muon coverage and
redesigned trigger and DAQ systems.
The new drift chamber (\abseta$<$1.0) and 
three silicon detectors (\abseta $<$1.8) along
with the new calorimeter in the end-cap region (``Plug'')
cover the whole pseudo-rapidity range up to \abseta = 3.6.

\section{$\wlnu$ and $\zll$ inclusive cross sections}

\begin{figure}
\bce
\epsfig{figure=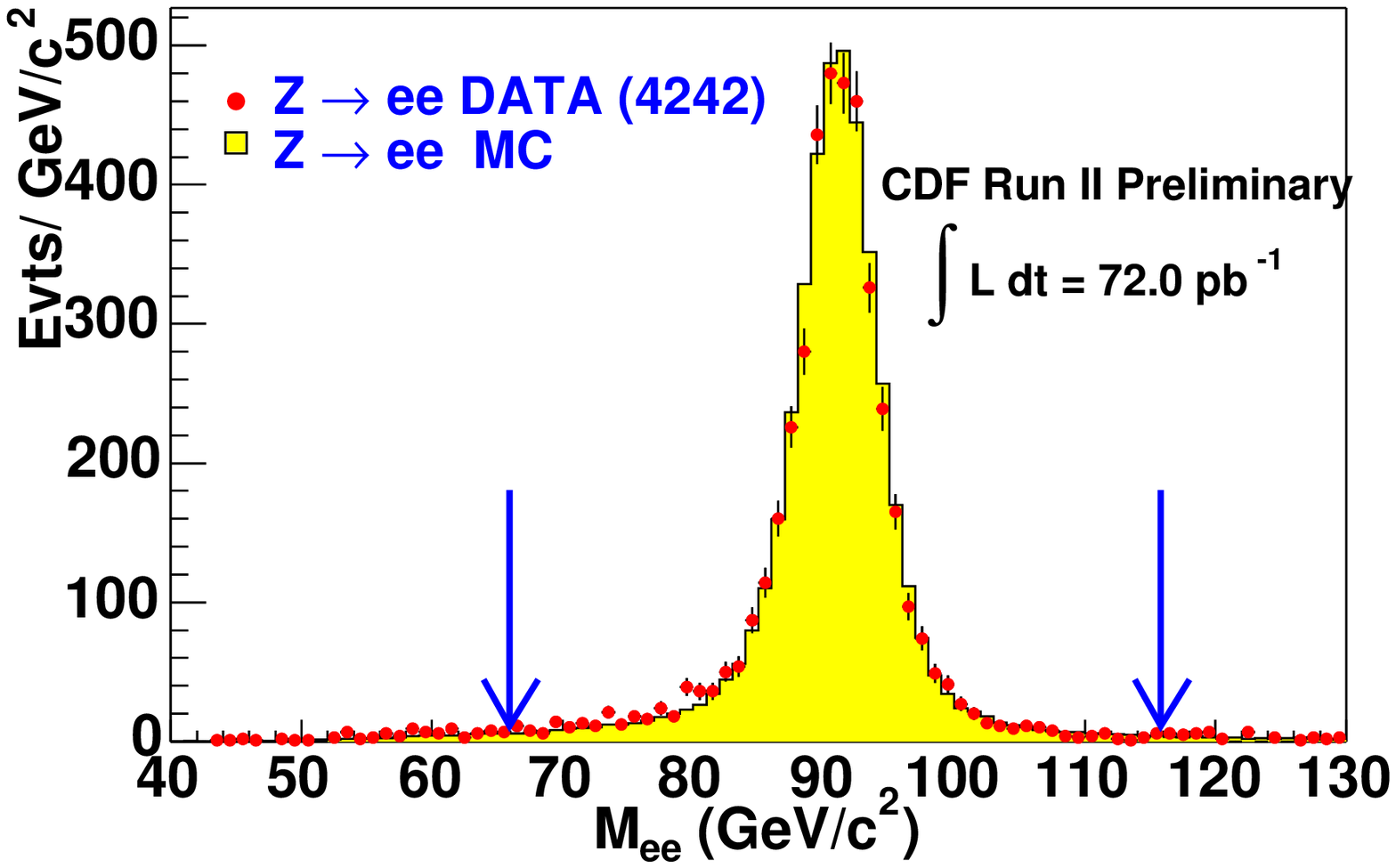,width=0.48\textwidth,height=7cm}
\epsfig{figure=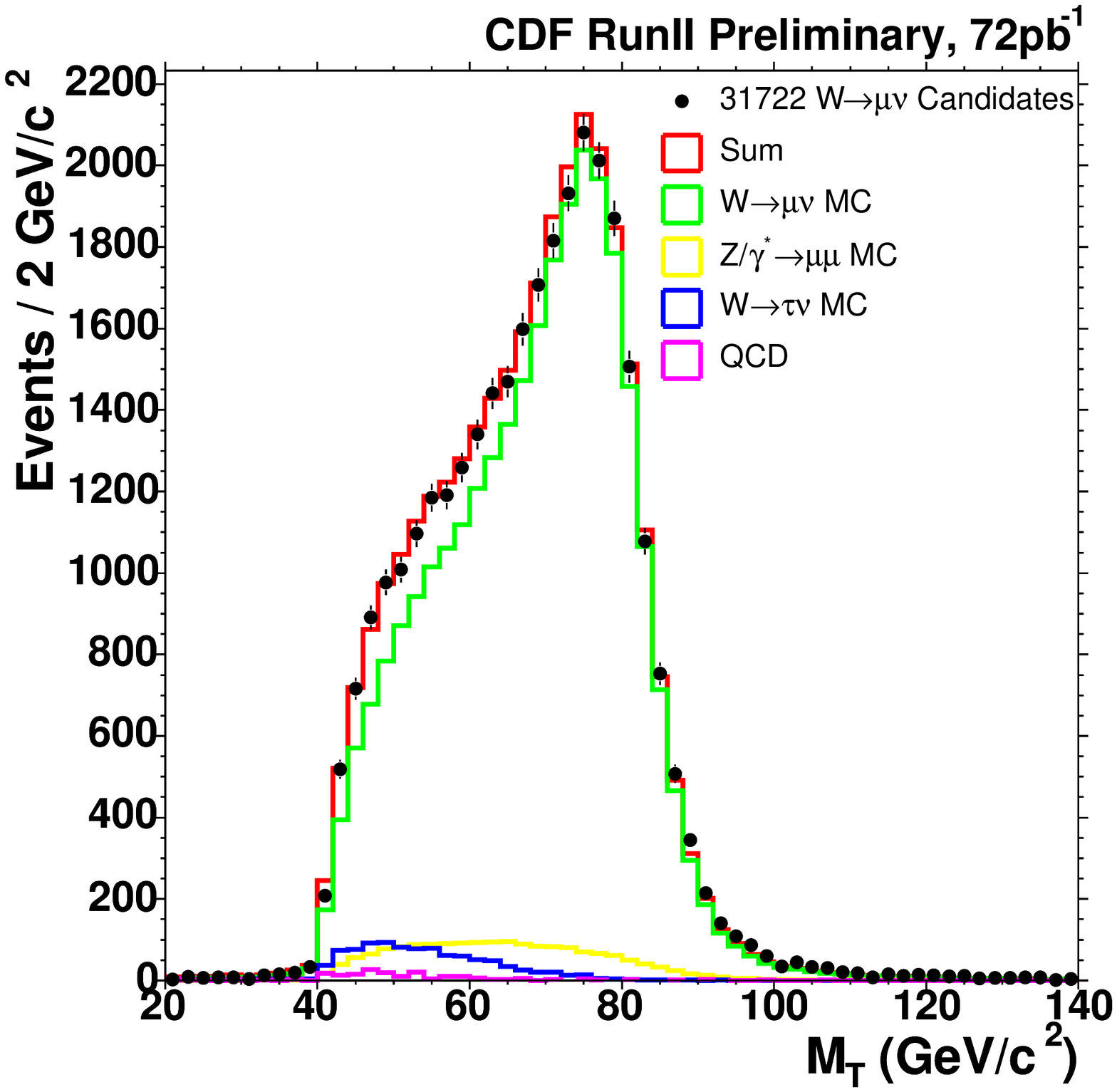,width=0.48\textwidth,height=7cm}
\ece
\caption{{\it On the left:} $\zee$ invariant mass distribution from CC and CP events;
	the arrows indicate the invariant mass window in which the measurement
	has been performed. {\it On the right:} $\wmnu$ transverse mass distribution from
 	data(dots), Monte Carlo signal and all considered sources of background(histograms).
\label{fig:masses}}
\end{figure}

\begin{figure}
\bce
\begin{tabular}{c c}
\epsfig{figure=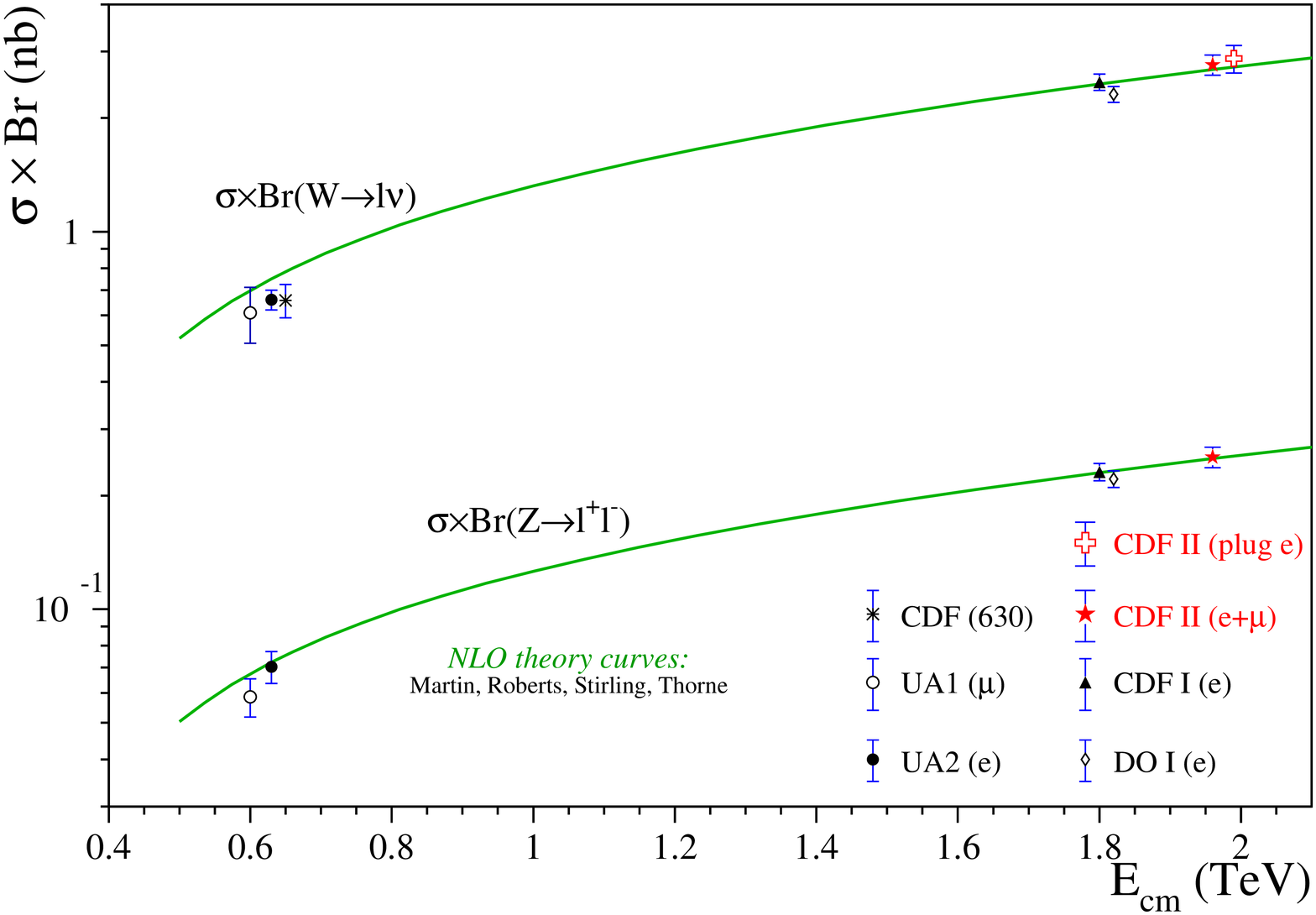,width=0.6\textwidth,height=7cm} &
%
\epsfig{figure=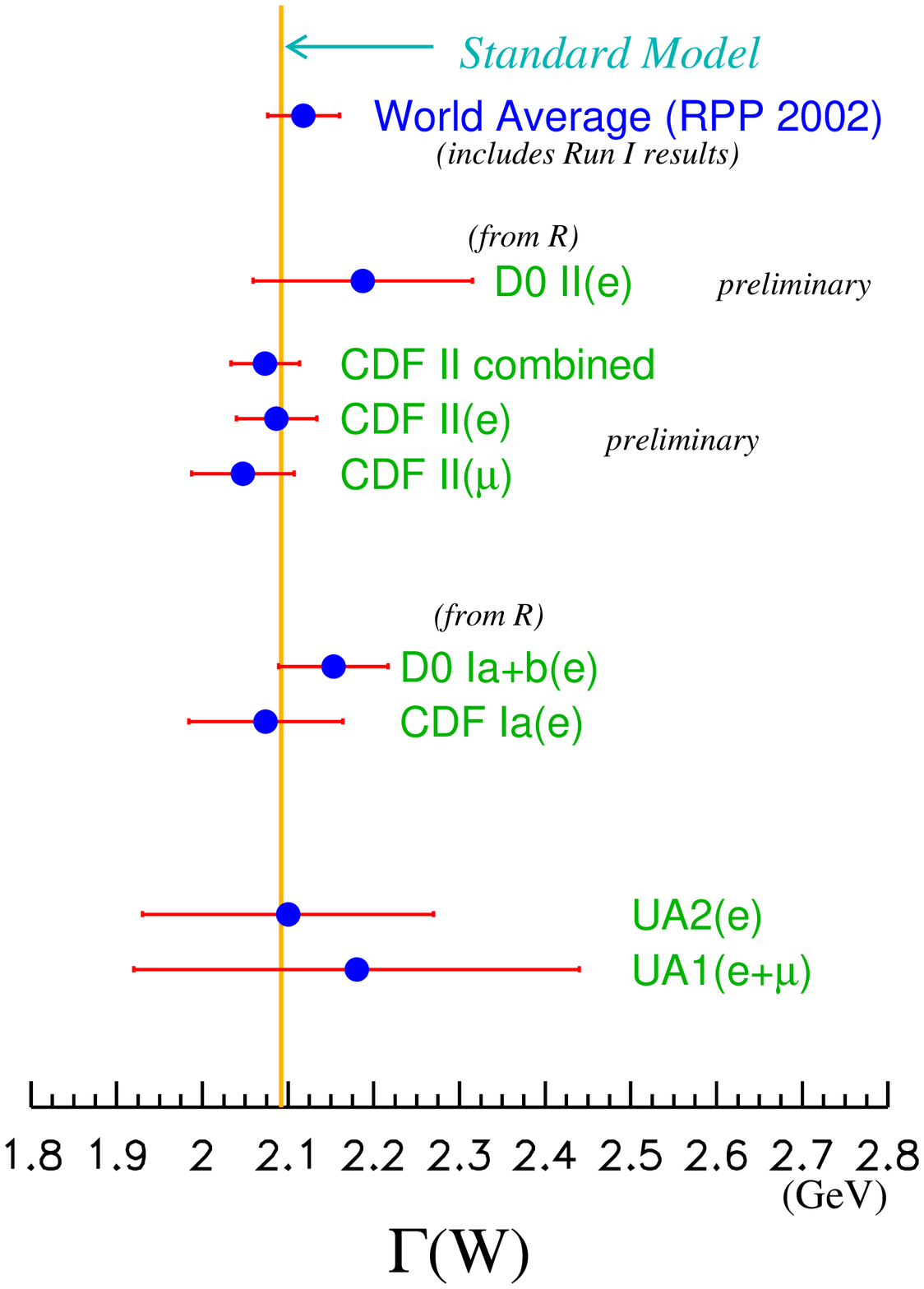,width=0.36\textwidth,height=7cm} \\
\end{tabular}
\ece
\caption{{\it On the left:} Recent $\zll$ and $\wlnu$ cross section measurements 
as a function of the center of mass energy for the CDF experiment, compared to 
other measurements in literature and to the NNLO calculation (as described in the text). 
{\it On the right:} Value of the total decay width of the $W$ boson extracted from the
measurement of the ratio of the $\wlnu$ to $\zll$ cross sections for the CDF and 
D\O\ experiments, compared to the Standard Model expectation (yellow band) and other
indirect measurements in literature.}\label{fig:gamma} \label{fig:xsecs}
\end{figure}

The $W$ and $Z$ cross section measurements,
with their high event
yields and relatively clean signatures, not only 
provide a solid test of the Standard Model 
but are extensively used in the calibration of the
calorimeter energy scales.
Furthermore, a precision measurement of the ratio of 
the cross sections
is sensitive to new physics states which might 
decay preferentially in either $W$ or $Z$.
In addition, confidence in these results
allows them to 
be used as normalization for successive measurements 
(such as the luminosity measurement and the top production 
cross section),
for which systematic and theoretical uncertainties would 
cancel in the ratio. Events are selected 
requiring one 
high-\pt\ isolated lepton and a second lepton 
(with slightly looser requirements) for the $Z$, 
or large missing transverse energy in the case of the $W$.
The results presented here include the 
measurements relative to the region of pseudo-rapidity
\abseta $<$ 1 and the new measurements of the $W$ and $Z$ 
cross sections performed using leptons reconstructed in the 
forward regions of the detector.
These include a new 
$W$ cross section measurement using electrons reconstructed in the
Plug calorimeter, and a combined $Z$ cross section,
where the first electron is found in the central region (\abseta $<$ 1.0)
and the second can be either in the central(CC) or in the Plug(CP)
(\abseta $<$ 2.8). The measurements in the muon channel include
the muons  reconstructed in the Central Muon Extension (CMX) 
sub-detector which extends the coverage from \abseta $<$ 0.6
up to \abseta $<$ 1. 
One of the most interesting new results of Run II at CDF is
the $\wtnu$ cross section measurement,
which uses the dedicated trigger selecting events through
the hadronic decay of the tau in association with 
large missing energy.
Although all these measurements still use the first 72 $\pbinv$ 
of data, time has been spent to obtain a thorough understanding of the 
systematic effects involved, which already limited the measurements
one year ago~\cite{mor2003}. Thanks to this effort the systematic
uncertainty has been reduced by one 
percentual point compared to the previous
result.
%
%
%
Distributions of the $Z$ invariant mass and the $W$ transverse mass 
in the electron  and muon channels respectively are shown in 
Figure~\ref{fig:masses}.
The event yield for each channel, background fraction,
acceptance times  efficiency and measured cross-section are shown in 
Table~\ref{tab:wzxsecs}; the increase in acceptance due to the extended
coverage in pseudo-rapidity can be observed~\cite{mor2003}.
	\begin{table}[t]
	\caption{The raw event yields and resulting inclusive $W,Z$ 
	cross sections measured
 	by the CDF experiment.  \label{tab:wzxsecs}}
	\bce \resizebox{\textwidth}{!} {
	\begin{tabular}{c c c c c c c}
	\hline
	{\bf Channel(mode)} &$\mathbf{ \eta_{lepton}}$ {\bf coverage} & $\mathcal{L}$ & {\bf Raw yield} & {\bf Background} 
	& {\bf A$\cdot \epsilon$} & $\mathrm{\mathbf{\sigma\cdot Br}}$ \\ 
	\hline
	$\wenu$   & \abseta$<$1    & 72  & 37574 & 4.4\% & 17.9\% & $2782\pm14_{stat}$$^{+61}_{-56\,syst}\pm167_{lum}$ pb\\
	$\wenu$&1.1$<$\abseta$<$2.8& 64  & 10461 & 8.7\% &  5.2\% & $2874\pm34_{stat}\pm67_{\,syst}\pm172_{lum}$ pb\\
	$\wmnu$   & \abseta$<$1    & 72  & 37574 &10.6\% & 14.4\% & $2772\pm16_{stat}$$^{+64}_{-60\,syst}\pm166_{lum}$ pb\\
	$\wtnu$   & \abseta$<$1    & 72  &  2345 &26.0\% &  0.9\% & $2620\pm70_{stat}\pm210_{syst}\pm160_{lum}$ pb\\
	$\zee$    & 1$^{st} e$:\abseta $<$ 1; 2$^{nd} e$:\abseta $<$ 2.8 & 
			        72  & 4242  & 1.5\% & 22.7\% & $255.2\pm3.9_{stat}$$^{+5.5}_{-5.4\,syst}\pm15.3_{lum}$ pb\\
	$\zmm$    &\abseta$<$1 &72  & 1785  & 0.7\% & 10.2\% & $248.9\pm5.9_{stat}$$^{+7.0}_{-6.2\,syst}\pm14.9_{lum}$ pb\\
	\hline
	\end{tabular} }
	\ece
	\end{table}
The combined measurements in the muon and electron channel~\footnote{
	The measurement of the $\wenu$ cross section using the electrons reconstructed 
	in the Plug calorimeter has not been included in the combination.}
are:
\begin{eqnarray*}
\sigma(\ppbar \rg Z^0 /\gamma^* \rg \ell \ell)  & = & 254.3\pm3.3_{stat}\pm4.3_{syst}\pm15.3_{lum}\,{\rm pb}\\
\sigma(\ppbar \rg W^\pm \rg \ell \nu) & = & 2777 \pm10_{stat}\pm52_{syst}\pm167_{lum}\, {\rm pb}.
\end{eqnarray*}
These measurements are in excellent agreement
with the NNLO theoretical calculations
at 1.96 \TeV~\cite{nnlo}
of $\sigma(\ppbar \rg Z^0 \rg \ell \ell)$ = 252$\pm$5 pb and 
$\sigma(\ppbar \rg W^\pm \rg \ell \nu)$ = 2690$\pm$50 pb, and previous 
measurements in literature~\cite{pdg}, as shown in 
Figure~\ref{fig:xsecs}(left). CDF and D\O\ results are compared elsewhere in 
these proceedings~\cite{marco}.
%
%
From the $\wtnu$ and $\wenu$ cross-section measurements 
CDF has extracted 
the ratio of the coupling constants  
$g^W_{\tau}/ g^W_e = 0.99\pm0.02_{stat}\pm0.04_{syst}$, probing
lepton universality. 
The signal ${\rm Z}^0\rightarrow \tau_{lep} \tau_{had}$ has been 
observed; the major challenge to the upcoming measurement of
the cross section is the study of the background, largely dominated 
by QCD di-jet events. 
%
%
Once finalized, this channel will be the ideal starting
point for all analyses including taus, particularly 
searches for Supersymmetry in models with high values of
tan$\beta$,    for  which the branching ratio 
$A/h \rg \tau\tau$ is enhanced~\cite{amy}.
From 
the ratio of the cross sections 
$\sigma(\ppbar \rg W^\pm \rg \ell \nu)$ to
$\sigma(\ppbar \rg Z^0 \rg \ell \ell)$ CDF has 
extracted the value of the total decay width of the $W$ boson,
which has
 been found to be: $\Gamma ({W} \rightarrow \ell \nu)$ = 
2071 $\pm$ 40 \MeV, where the value is the combined value in 
the electron and muon channels
and the uncertainty includes the statistical and
systematic contributions. 
This value is consistent with 
the LEP direct measurement of 2.150$\pm$0.091 \GeV, the Run I CDF and D\O\ combined 
measurement of 2.115 $\pm$ 0.105 \GeV, and the D\O\ Run II measurement of 
2187$\pm$128 \MeV, obtained with 42 $\pbinv$ of data in the electron channel~\cite{marco}.
This value alone is already as precise as the current world average value of 
2092$\pm$40 \MeV, and it is the most precise experimental measurement of
$\Gamma(W)$ up to now.
In Figure~\ref{fig:gamma}(right) the CDF single and combined values are compared 
to the Standard Model expectation,  other indirect measurements and the
D\O\ Run II measurement.  From the ratios measured separately in the 
electron and muon channels, CDF has also extracted the 
 ratio of the coupling constants  
$g_{\mu}/ g_e = 1.011\pm0.018_{stat+syst}$, another probe of 
lepton universality.

\section{$Z^0$ forward-backward asymmetry}

\begin{figure} \bce
\epsfig{figure=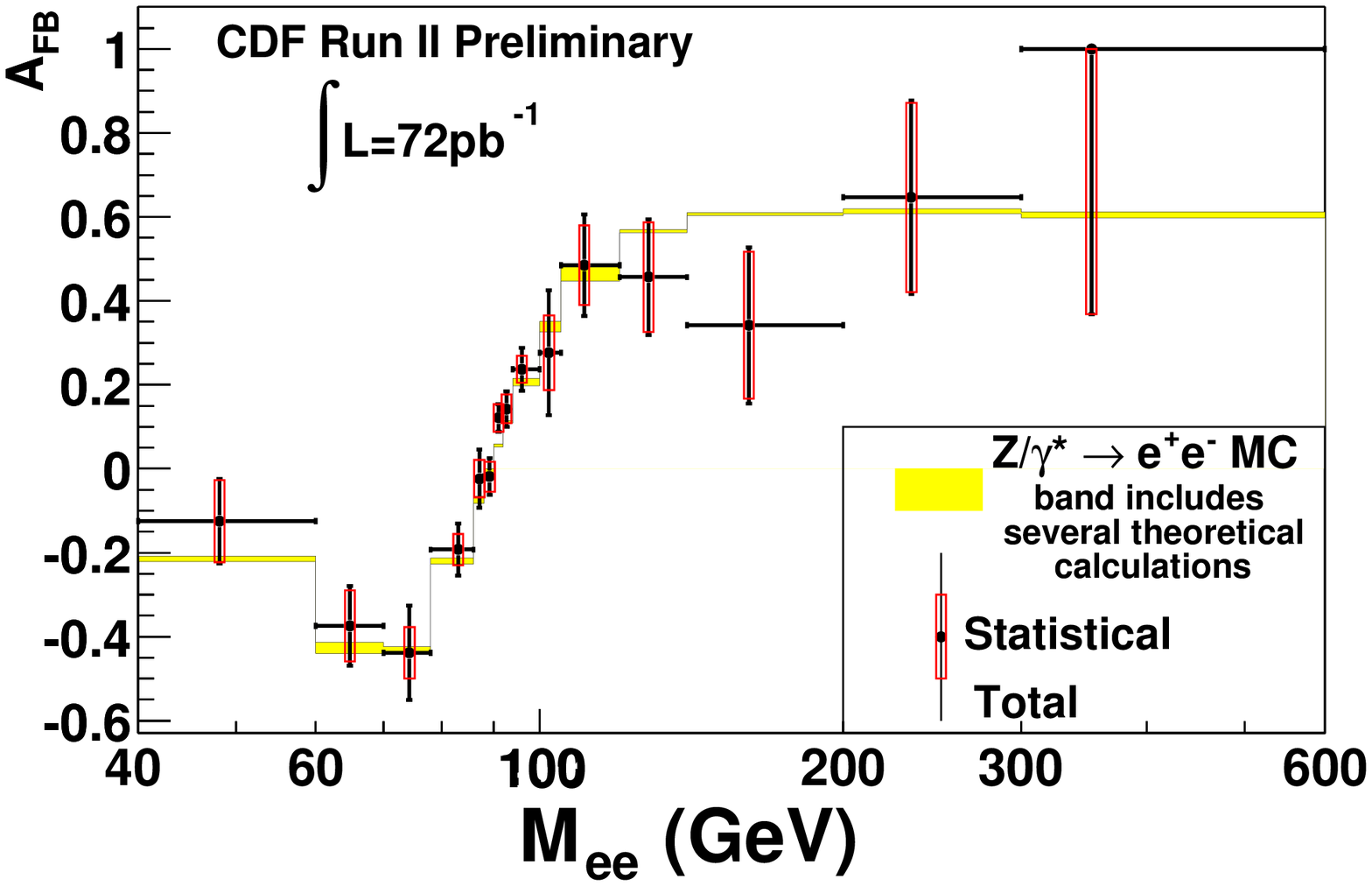,width=0.48\textwidth,height=7cm}
\epsfig{figure=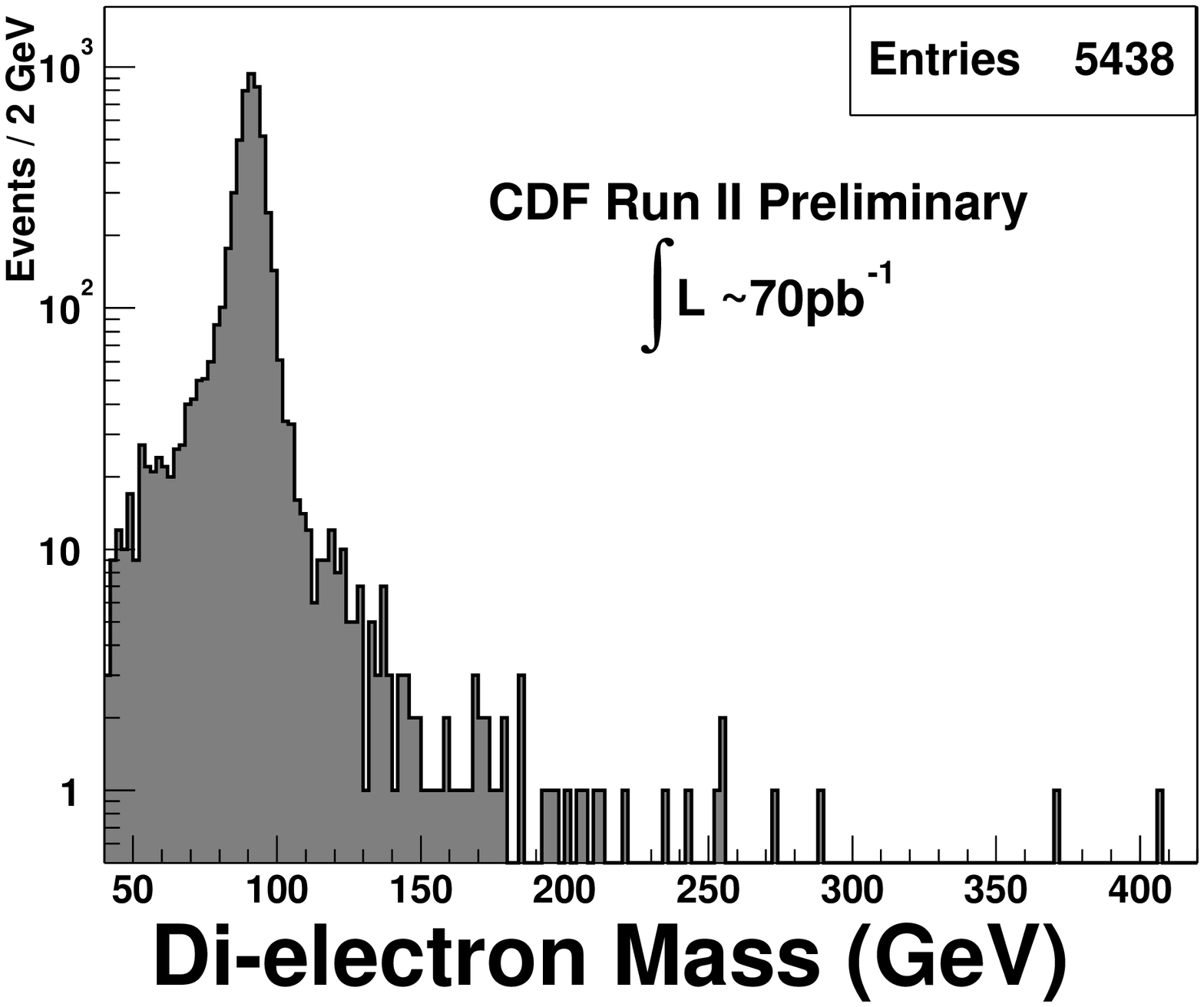,width=0.48\textwidth,height=7cm}
\ece
\caption{
{\it On the left:} CDF measurement of the forward-backward asymmetry in the 
di-electron channel for the entire electron-positron invariant mass spectrum, 
compared to the theory predictions.
{\it On the right:} Invariant mass spectrum of the $Z$ candidate events used in the 
measurement of  $A_{fb}$.
}\label{fig:afb}
\end{figure}

A forward-backward asymmetry can be observed in the decay of the leptons
in the process 
$\bar{q}q\rg Z/\gamma^* \rg \ell\ell$, due to the presence of both 
vector and axial-vector couplings of electroweak bosons to fermions.
The measurement of $A_{fb}$, defined as
$A_{fb} = \frac{N_F - N_B }{N_F + N_B}$, 
where $N_F$ and $N_B$ are the number of forward and backward
events (in the rest frame of the lepton pair) respectively, is 
a direct probe of the strengths of 
the couplings involved. Furthermore, a measurement of $A_{fb}$ is 
sensitive to the presence of high mass Drell-Yan(DY) resonances
outside the Standard Model~\cite{ming}.
The CDF measurement of $A_{fb}$ in 72.0 $\pbinv$ 
(shown in 
Figure~\ref{fig:afb}) is one of the first analyses in Run II to make
use of the full coverage $\abseta < 3.0$  
provided by the new Plug calorimeter.
The selection of the events requires a high-\pt\ electron
in the central region and a second one anywhere in the
detector, with the main background consisting of di-jet QCD
events.
While the LEP measurement~\cite{lep} is better in the
low invariant mass region, the measurement for M$_{ee} >$ 200 \GeV\
is unique at the Tevatron. This measurement is currently
being updated with the full 200 $\pbinv$ of data.

\section{Di-boson results}

\begin{figure} \bce
\epsfig{figure=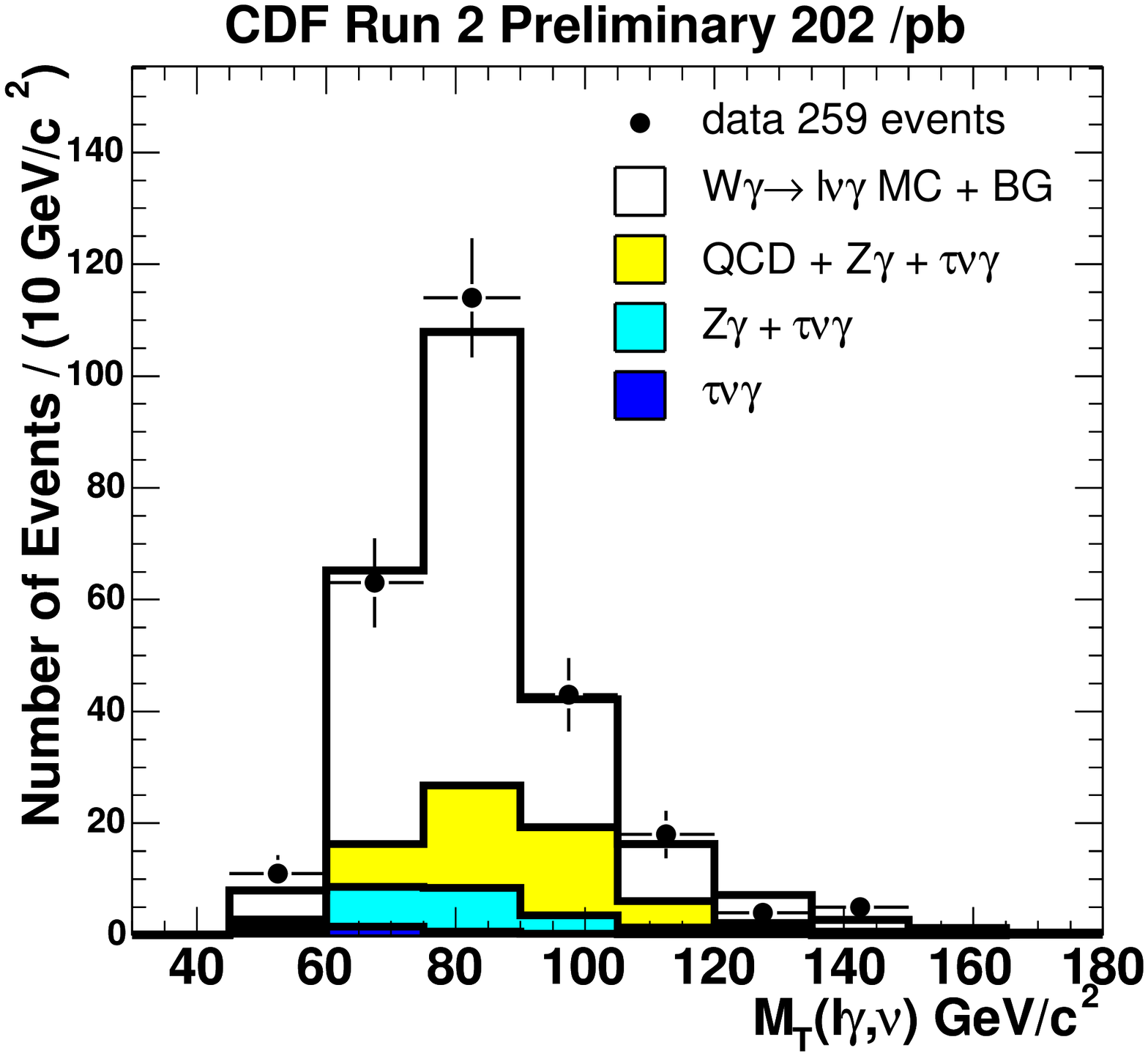, width=0.48\textwidth,height=7cm}
\epsfig{figure=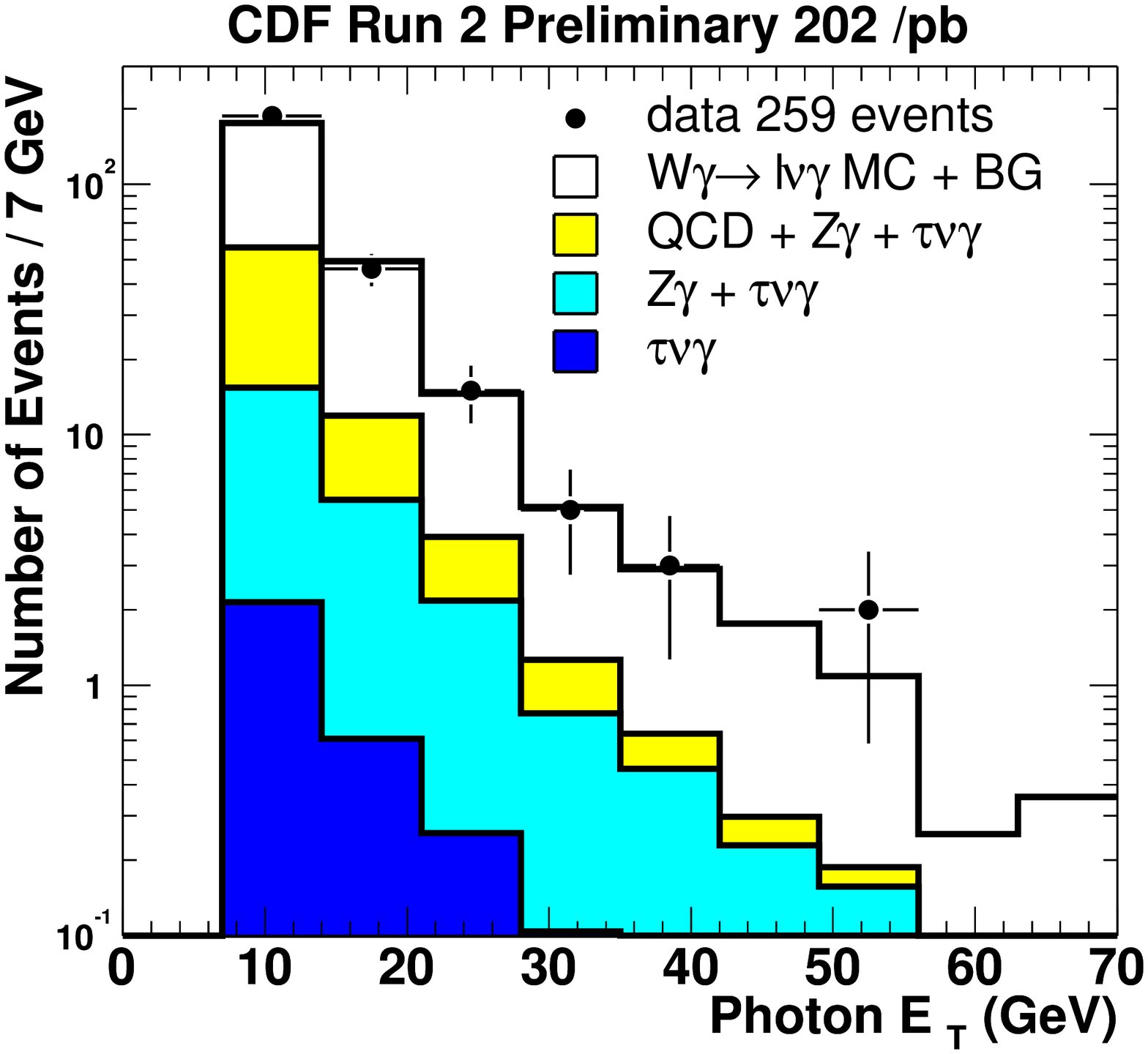,width=0.48\textwidth,height=7cm}
\ece
\caption{{\it On the left:} Transverse mass distribution of the lepton, photon and neutrino
	for $W\gamma$ events. {\it On the right:} Photon $E_T$ distribution of
	$W\gamma$ events on a 
	logarithmic scale for the MC signal (open histogram), data (dots) and
	all the expected sources of background. Presence of anomalous coupling would 
	modify the photon spectrum towards higher values of $E_T$.} \label{fig:wgamma}
\end{figure}

\begin{figure} \bce
\epsfig{figure=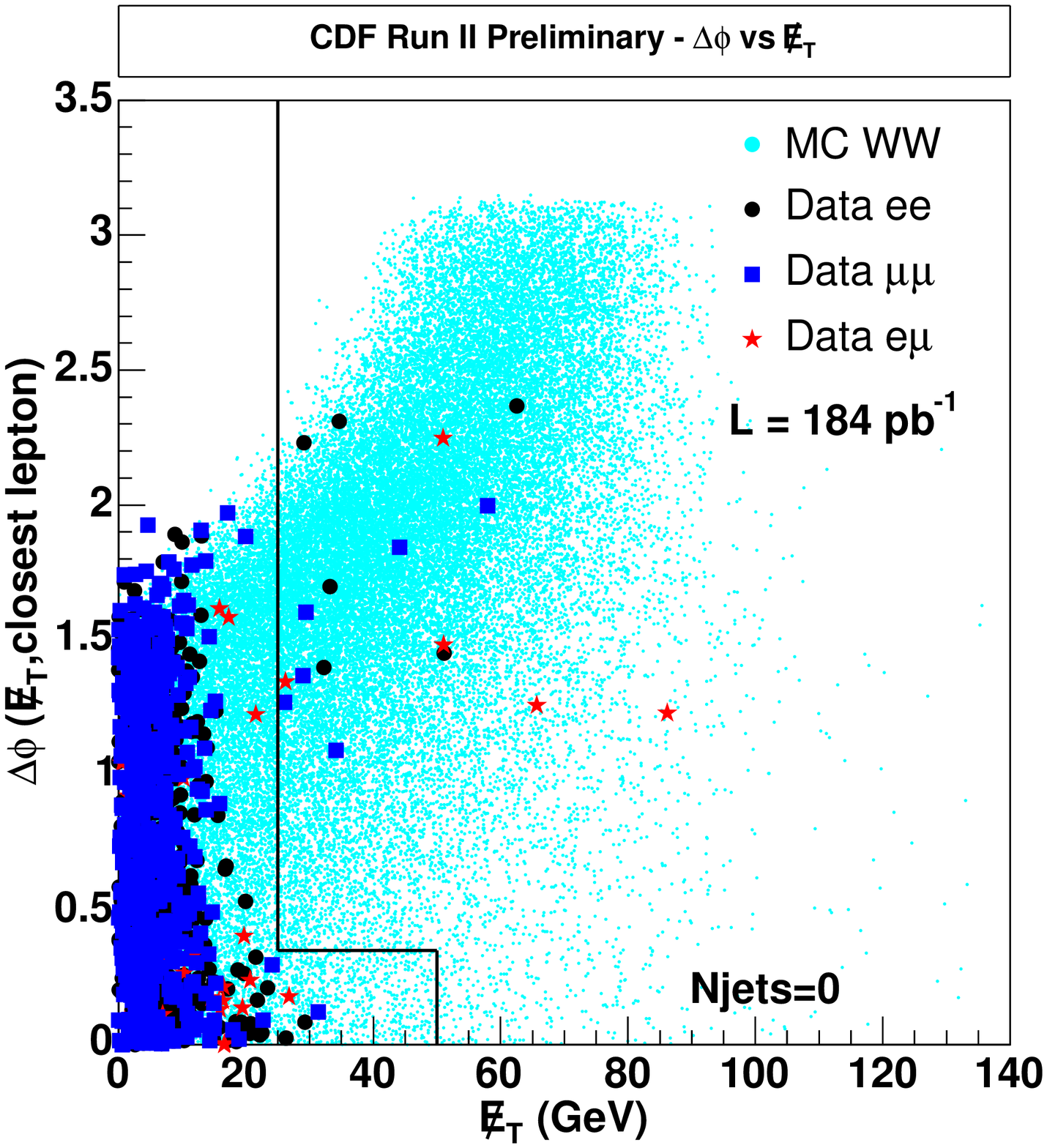,width=0.48\textwidth,height=7cm}
\epsfig{figure=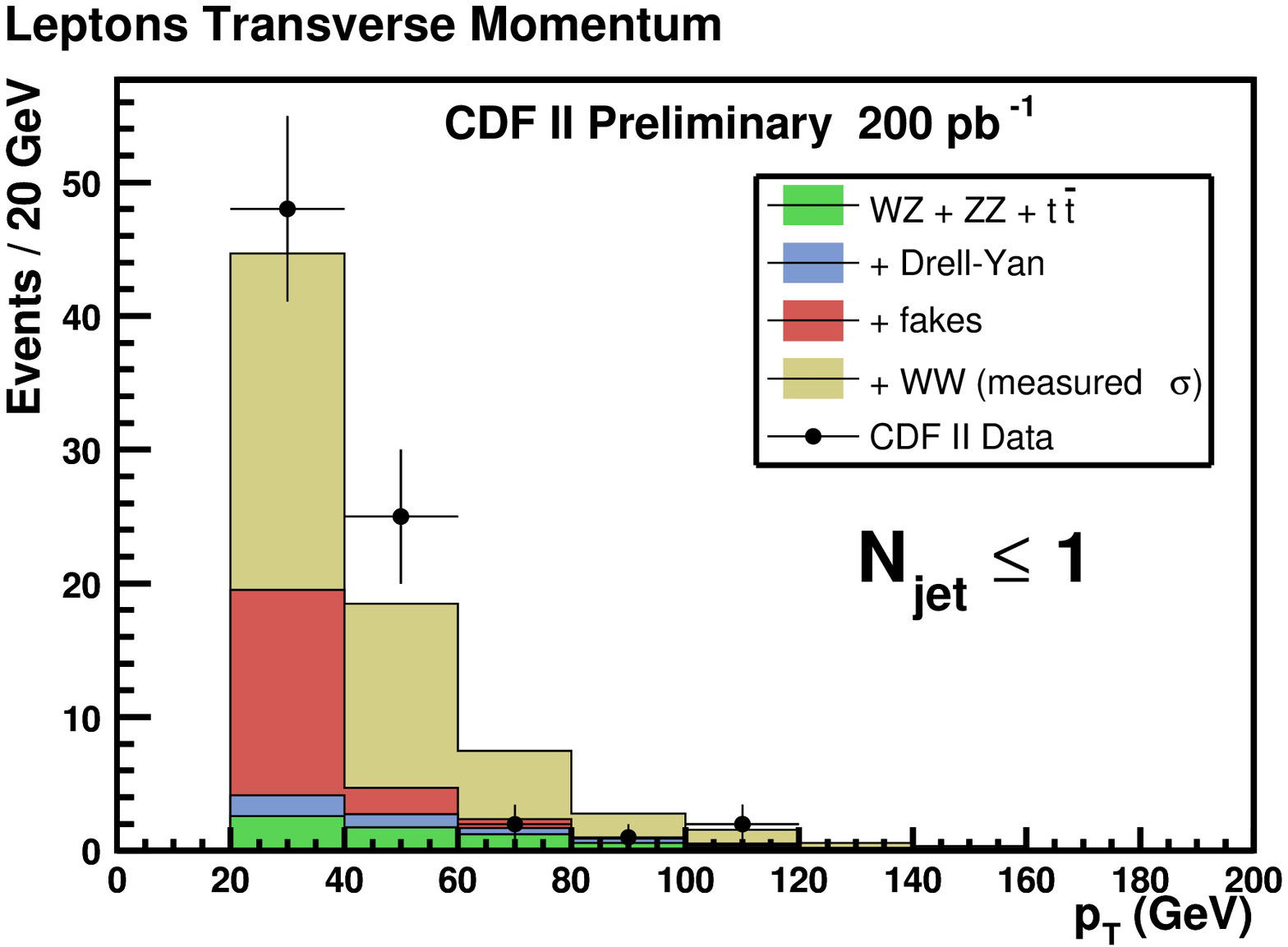,width=0.48\textwidth,height=7cm}
\ece
\caption{$\Delta\phi(\met, \ell)$ as a function of \met\ for the 17 $WW$ candidate 
	 events from the ``di-lepton'' analysis (left), and the $p_T$ distribution 
	of the two lepton tracks of the 39 ``lepton+track'' $WW$ events (right). In the 
	latter all the sources of expected background are also indicated.}\label{fig:ww}
\end{figure}

Associated $W\gamma$ and $Z\gamma$ production is an important
test of the non-Abelian nature of the SM as it is sensitive to 
triple-gauge boson interactions and thus to physics beyond
the Standard Model.
Using the same 
selection as in the $W$ and $Z$ inclusive cross section measurements but 
with the addition of a high-energy ($E_T^\gamma > 7$ \GeV)
photon isolated from the lepton ($\Delta R(\gamma - \ell) > 0.7$), 
CDF measured the cross sections 
$\sigma(\ppbar \rg W\gamma)\cdot Br(W\rg \ell\nu)$ and  
$\sigma(\ppbar \rg Z\gamma)\cdot Br(Z\rg \ell\ell)$ in 202 $\pbinv$ of Run II data.
The values along with the number of events expected and observed in the data
are shown in Table~\ref{tab:wzgamma}.
	\begin{table}[htb]
	\caption{The raw event yields and resulting $W,Z\gamma$ cross sections measured
 	by CDF in 202 $\pbinv$ of data.
	}\label{tab:wzgamma} 
	\centering{ \resizebox{\textwidth}{!} { 
	\begin{tabular}{c c c c} 
	\hline
	{\bf Channel(mode)} & {\bf N events expected } & {\bf N events observed } & $\mathrm{\mathbf{\sigma\cdot Br}}$ \\ 
	                    & {\bf (signal MC + SM background)} & {\bf in the data} &                        \\
	\hline
$W^\pm\gamma\rg{\ell}^\pm \nu\gamma$ 
        & $255.63\pm2.13_{stat}\pm26.43_{syst}$ & 259 & $19.7 \pm 1.7_{stat}\pm 2.0_{syst}\pm 1.2_{lum}$ pb\\
$Z^0\gamma\rg{\ell}^+ {\ell}^-\gamma$
        & $70.46\pm4.00_{syst}^{(*)}$ & 69 & $5.3\pm0.6_{stat}\pm 0.3_{syst}\pm 0.3_{lum}$ pb\\
	\hline
	$^{(*)}$~{\it The statistical uncertainty from the MC samples is negligible.}
	\end{tabular} }}
	\end{table}
All the measurements are consistent with the SM predictions~\cite{wgrad,wzgamma} of
$\sigma(\ppbar \rg W\gamma)\cdot Br(W\rg \ell\nu) = 19.3 \pm 1.3$ pb and 
$\sigma(\ppbar \rg Z\gamma)\cdot Br(Z\rg \ell\ell) = 5.4 \pm 0.4$ pb.
Figure~\ref{fig:wgamma}(left) shows the transverse mass distribution 
reconstructed using the lepton, photon and missing energy information,
and the transverse energy of the photon in $W\gamma$ events (right).
As the presence 
of Anomalous Gauge Couplings (AGC) in the three bosons vertex
would modify the latter distribution 
boosting the photon towards higher $E_{T}$~\cite{agc}, the photon spectrum is a 
direct test of the presence of AGC. Preliminary results 
on Anomalous Gauge Coupling extraction from the photon $E_T$ spectrum
 are expected by the end of 2004.\\
Heavy di-boson measurements such as $WW, WZ$ and $ZZ$ also lead to new 
limits on triple gauge boson 
couplings and are important background to $t\bar{t}\rg \ell\ell$+jets,
Higgs and exotic signatures.
CDF has looked for $WW$ events in two separate complementary ways.
One analysis has been optimised for signal over background ratio, selecting
events with two isolated leptons (according to similar criteria
as the ones used in the inclusive cross section analyses), large 
missing energy and the absence of jets. 
The second analysis applies looser selection criteria on the second lepton (which
can also be just a high-$p_T$ track, allowing the presence of electrons, 
muons and taus), with the aim of improving the acceptance of the signal.
The number of observed and expected events are shown in Table~\ref{tab:ww},
along with the cross section results for the first 
(``di-lepton'') and second (``lepton+track'') method.
The results are consistent with each other and 
with the theoretical calculation of 12.5$\pm$0.8 pb~\cite{ww}.
\begin{table}[htb]
\caption{The raw event yields and resulting $WW$ cross sections measured by 
	the ``di-lepton'' and ``lepton+track'' analyses in CDF with 202 
	$\pbinv$ of data.}  \label{tab:ww}  
\centering{ \resizebox{\textwidth}{!} { 
\begin{tabular}{c c c c c} 
\hline
{\bf Method}&{\bf N events expected }&{\bf N events observed }& {\bf S/B}& $\mathbf \sigma(p\bar{p}\ra WW)$\\ 
          &{\bf (signal MC + SM background)}&{\bf in the data}       & &         \\
\hline
di-lepton    & $16.1\pm1.6_{stat+syst}$ & 17 & 2.3 &
$14.3^{+5.6}_{-4.9\,stat} \pm 1.6_{syst}\pm 0.9_{lum}$ pb\\
lepton+track & $31.5\pm1.0_{stat}$ & 39 & 1.1 &
$19.4\pm5.1_{stat}\pm 3.5_{syst}\pm 1.2_{lum}$ pb\\
\hline \end{tabular} } }
\end{table}
Figure~\ref{fig:ww} shows the kinematic
distribution of the ``di-lepton'' events compared with the Monte Carlo 
prediction (left), and the $p_T$ distribution of the leptons
in the ``lepton+track'' sample (right).

\section{Towards the $W$ mass}

\begin{figure}
\bce
\epsfig{figure=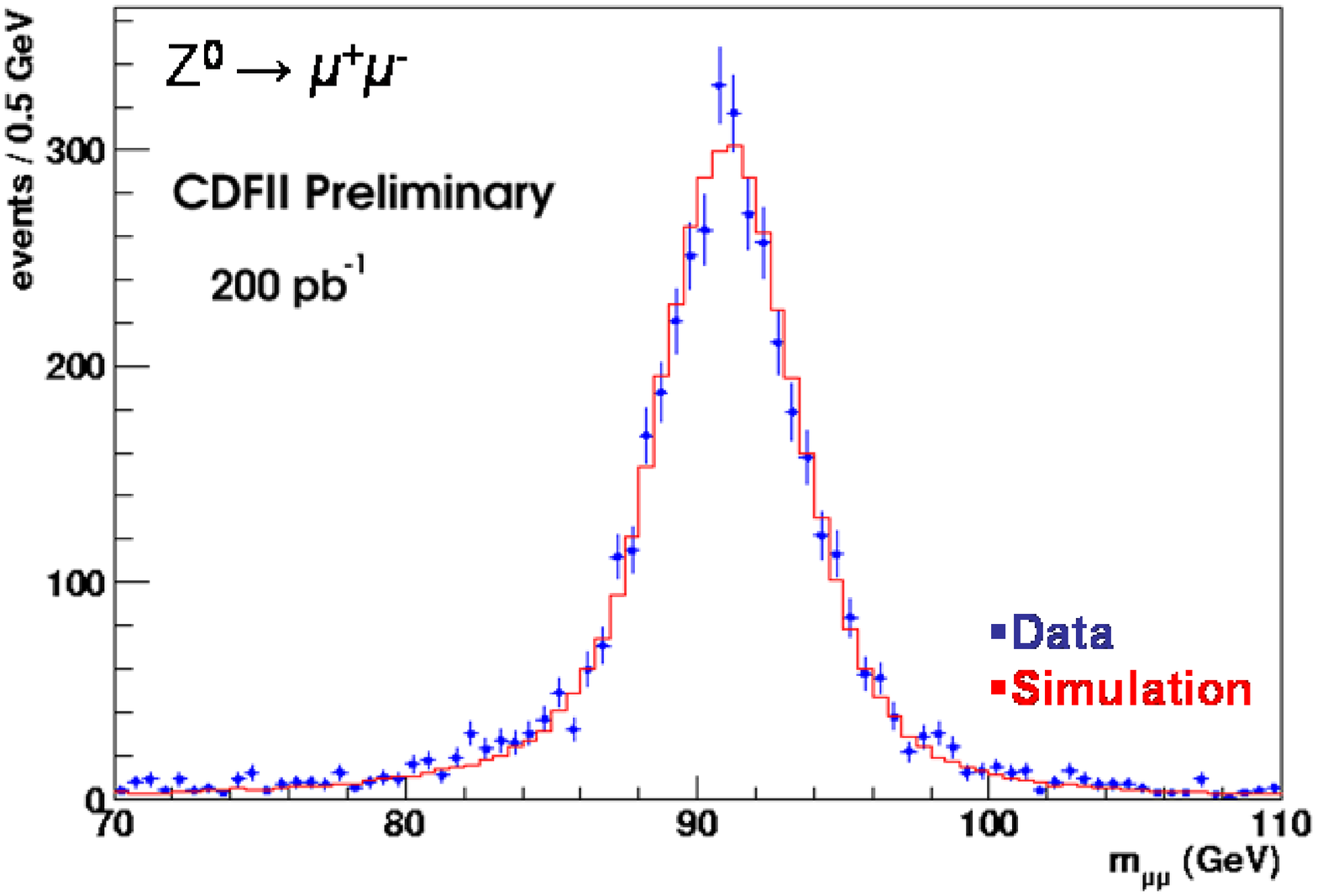,width=0.48\textwidth,height=7.0cm}
\epsfig{figure=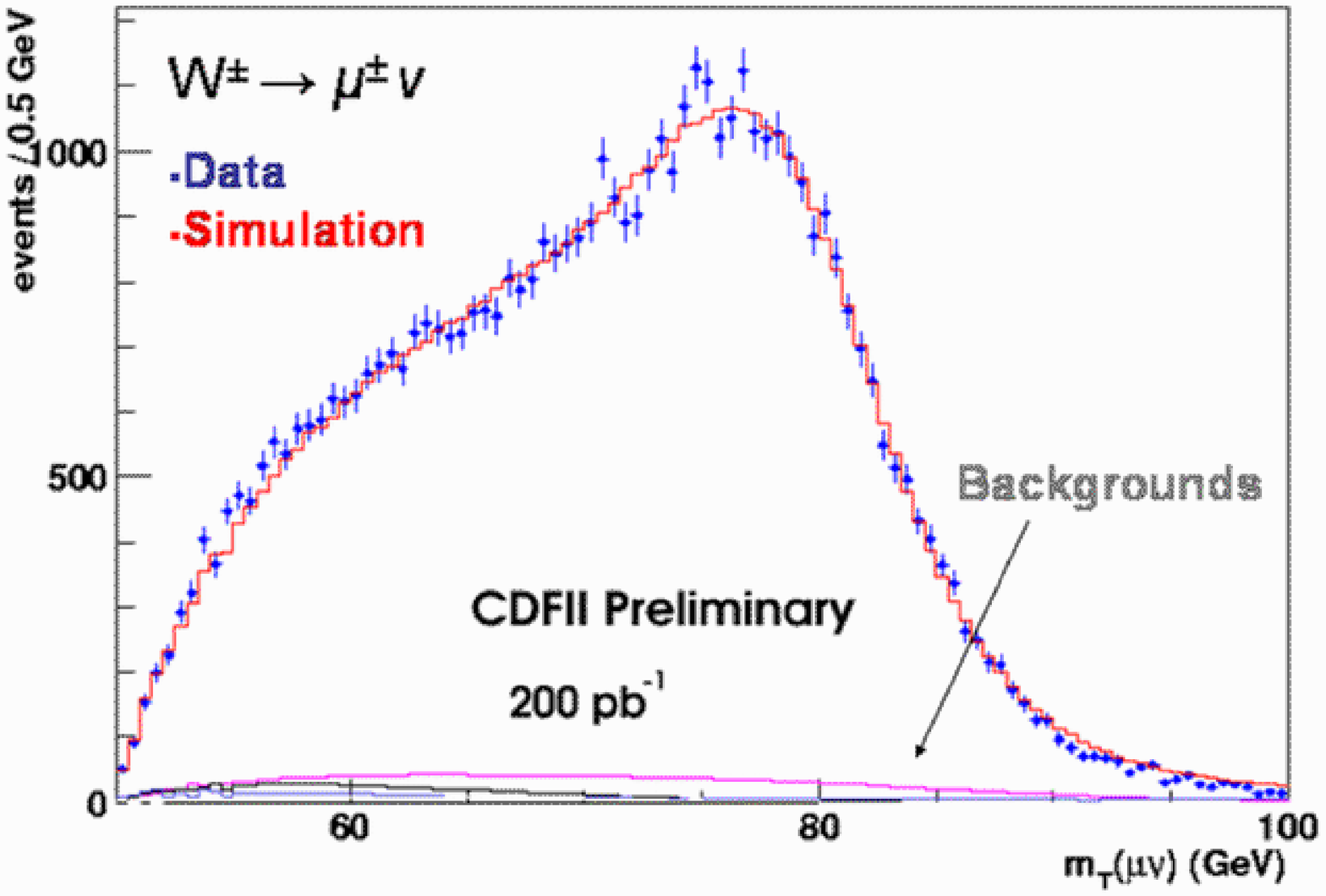,width=0.48\textwidth,height=7.0cm}
\ece
\caption{{\it On the left:} $\zmm$ invariant mass distribution 
	from data (dots) and simulation (histogram).
	{\it On the right:} $\wmnu$ transverse mass distribution 
	from  data (dots) and simulation (histogram).
	In the latter the contributions from all sources 
	of background are also shown.
\label{fig:wmass}}
\end{figure}

The mass of the $W$ boson is one of the most important parameters of the 
Standard Model, as its precise knowledge, combined with that of the
top mass, results in a significant constraint on the mass of the as
yet unobserved Higgs particle. From the experimental point of view this
is a challenging measurement as it requires a precise knowledge of 
the detector and the best understanding of its performance.
The method adopted by CDF consists in using Monte 
Carlo templates with different values of the $W$ mass to fit several
distributions of signal plus total expected background, and extract the
value of M$_W$ from the distribution which best matches the data.
Work is already in progress at CDF in understanding the key 
components relevant to this measurement: the track momentum scale 
and the missing energy resolution.
Figure~\ref{fig:wmass}(left) compares the $\zmm$ invariant mass 
distribution in data and Monte Carlo signal, generated with 
RESBOS~\cite{resbos} and simulated
using the best knowledge of the detector. This distribution will be used 
to test the linearity of the momentum scale, which will 
be measured using $J/\Psi$ events.
The $W$ transverse mass distribution, also shown in 
Figure~\ref{fig:wmass}(right), is directly affected by the resolution of
the missing energy measurement, representing the energy of
the escaping neutrino.
CDF and D\O\ have recently published~\cite{wmass} the Run~I
combined result of $M(W) = 80.456 \pm 0.059$ \GeV, 
with a statistical uncertainty of approximately 40 \MeV.
The goal of the Run II analysis is to use
data to control the systematic uncertainties, such that they scale with
the statistics of the data.
%
From the tail of the $W$ transverse mass distribution a direct measurement 
of $\Gamma(W)$ can be performed. As most of the systematics effects 
important in the extraction of the width are the same studied for the 
$W$ mass measurement, a preliminary result of the direct measurement of 
$\Gamma(W)$ will naturally follow the $W$ mass measurement expected
by the summer 2004.

\section{Conclusions and prospects}

Run II at the Tevatron is well under way.
The CDF detector is fully commissioned
and is taking high-quality physics data.
The Electroweak physics programme 
has reestablished 
the basic measurements, benchmarks for the understanding of the
detector backgrounds and lepton identification,
and the precision measurements 
are already competitive with the Run I results.
We expect to finalize results on di-boson production 
processes 
and differential cross  sections for summer 2004, along with 
a preliminary measurement on the $W$ charge asymmetry and $W$ mass.


\section*{References}
\begin{thebibliography}{99}

\bibitem{peter} P. Garbincius, ``Tevatron performance'', {\it These proceedings}.

\bibitem{top}
M. Cacciari, S. Frixione, M. L. Mangano, P. Nason and G. Ridolfi,
``The t anti-t cross-section at 1.8-TeV and 1.96-TeV: A study of the  systematics due to parton densities and scale dependence,'' hep-ph/0303085.

\bibitem{ken} K. Bloom, ``Top Physics at CDF'', {\it These proceedings};\\
Y. Kulik, Top Physics at D\O, {\it These proceedings}.

\bibitem{mor2003}
P. Koehn, ``Top, $W$ and $Z$ results at CDF'',  XXXVIII Rencontres de Moriond
Electroweak Interactions and Unified Theories Les Arcs, 15-22$^{nd}$March 2003.

\bibitem{nnlo}
A.D. Martin {\it et al.} , \Journal{\PLB}{531}{216}{2002};
W.J. Stirling, {\it private communication}.

\bibitem{pdg}
K. Hagiwara {\it et. al.}, \Journal{\PRD}{66}{010001}{2002}.

\bibitem{marco} M. Verzocchi, ``Electroweak Physics results at D\O'', {\it These proceedings}.

\bibitem{amy}
A. Connolly (CDF Collaboration), hep-ex/0212016.

\bibitem{ming} S.M. Wang, ``Exotic searches at CDF and D\O'', {\it These proceedings}.

\bibitem{lep}
The LEP collaborations:{ALEPH,DELPHI,L3} and OPAL, 
hep-ex/0101027 (2000).

\bibitem{wgrad}
U. Baur, S. Keller and D. Wackeroth,
``Electroweak radiative corrections to $W$ boson production in hadronic
collisions,'' \Journal{\PRD}{59}{013002}{1999}.

\bibitem{wzgamma} U. Baur, {\it private communication}.

\bibitem{agc}
H.~Aihara {\it et al.},
``Anomalous gauge boson interactions'', hep-ph/9503425.

\bibitem{ww}J.M. Campbell and R.K. Ellis, \Journal{\PRD}{60}{113006}{1999}.

\bibitem{resbos}
C.~Balazs and C.~P.~Yuan,
``Soft gluon effects on lepton pairs at hadron colliders'',
\Journal{\PRD}{56}{5558}{1997}.

\bibitem{wmass} The CDF and D\O\ Collaborations,
``Combination of CDF and D0 results on $W$ boson mass and width,'' 
hep-ex/0311039 (2003).

\bibitem{cdfwmass}
T. Affolder {\it et al.} [CDF Collaboration],
``Measurement of the $W$ boson mass with the Collider Detector at Fermilab,''
\Journal{\PRD}{64}{052001}{2001}.

\end{thebibliography}
\end{document}